\documentclass[a4paper,10pt, twocolumn]{article} 

\usepackage[a4paper, margin=0.8in]{geometry} 
\usepackage{graphicx} 
\usepackage{amsmath, amssymb} 
\usepackage{hyperref} 
\usepackage{booktabs} 
\usepackage{caption} 
\usepackage{setspace} 
\usepackage{titlesec} 
\usepackage{listings} 
\usepackage{xcolor} 
\usepackage{titling}

\titleformat{\section}{\large\bfseries}{\thesection}{1em}{}
\titleformat{\subsection}{\normalsize\bfseries}{\thesubsection}{1em}{}

\definecolor{dkgreen}{rgb}{0,0.6,0}
\definecolor{gray}{rgb}{0.5,0.5,0.5}
\definecolor{mauve}{rgb}{0.58,0,0.82}
\definecolor{eclipseStrings}{RGB}{42,0.0,255}
\definecolor{eclipseKeywords}{RGB}{127,0,85}

\colorlet{numb}{magenta!60!black}

\lstdefinelanguage{json}{
	basicstyle=\normalfont\ttfamily,
	commentstyle=\color{eclipseStrings}, 
	stringstyle=\color{eclipseKeywords}, 
	numbers=left,
	numberstyle=\scriptsize,
	stepnumber=1,
	numbersep=8pt,
	showstringspaces=false,
	breaklines=true,
	frame=lines,
	string=[s]{"}{"},
	comment=[l]{:\ "},
	morecomment=[l]{:"},
	literate=
	*{0}{{{\color{numb}0}}}{1}
	{1}{{{\color{numb}1}}}{1}
	{2}{{{\color{numb}2}}}{1}
	{3}{{{\color{numb}3}}}{1}
	{4}{{{\color{numb}4}}}{1}
	{5}{{{\color{numb}5}}}{1}
	{6}{{{\color{numb}6}}}{1}
	{7}{{{\color{numb}7}}}{1}
	{8}{{{\color{numb}8}}}{1}
	{9}{{{\color{numb}9}}}{1}
}

\title{\textbf{Designing a Geo-Tourism App: A Principled Approach} }
\author{Ciuffoletti, Augusto \\ University of Pisa \\ \texttt{augusto.ciuffoletti@gmail.com}}
\date{March 31, 2025}

\begin{document}
	
	\maketitle 
	
	\let\thefootnote\relax\footnotetext{This research was conducted as part of the \href{https://sites.google.com/view/prin-underlandscape}{Underlandscape project}, a PRIN research initiative funded by the Italian Ministry of University and Research (MUR).}
	
	\begin{abstract}
		\noindent 
		Walking along trails in natural areas is a rewarding experience, but visitors sometimes need proper assistance to enhance their enjoyment, maximize learning, and ensure safety. Over the years, various signage techniques have been introduced, but today, the widespread use of smartphones offers new opportunities for visitor support.
		
		In this paper, we outline the key principles for designing an Android app tailored for geotourists. Our approach begins by defining user personas and deriving app requirements based on their needs. We then present a proof of concept that addresses the critical aspects identified during the design process.
	\end{abstract}
	
	\section{Introduction}
	Walking along trails in natural areas offers a unique and enriching experience. Visitors not only get to immerse themselves into nature but also have the opportunity to learn about the environment and its diverse features. However, to fully benefit from these experiences, proper support is essential to enhance enjoyment, maximize learning, and ensure visitor safety.
	
	Historically, signage systems have played a critical role in guiding visitors through wilderness areas. These traditional methods, while effective to some extent, are static and often fail to provide personalized or real-time information. With the widespread adoption of smartphones, there is now an unprecedented opportunity to transform the way visitors interact with natural spaces. Modern mobile devices offer dynamic capabilities such as real-time navigation, context-aware information delivery, and interactive content, all of which can significantly elevate the visitor experience.
	
	This paper explores the integration of smartphone technology into visitor support systems and follows a holistic approach to address the associated challenges. Our research not only focuses on the operational aspects of a guidance app but also takes into account the broader ecosystem of stakeholders. This includes the local community that benefits from the valorization of natural areas and is responsible for the maintenance of the system, as well as the dedicated team that develops the content. By considering these diverse perspectives, our framework aims to provide robust, real-time guidance while fostering a sustainable, community-driven model for long-term success.
	
	In the following sections, we begin by defining all aspects of the problem—including the technical challenges, the requirements of the visitors, and the responsibilities of the local community and content developers. Next, we detail our proposed solution, which integrates advanced smartphone technology with a holistic framework that addresses not only the operational functionality of the guidance app but also the needs and roles of all key stakeholders. Finally, we discuss how our approach relates to and extends existing works, highlighting its unique contributions to the field.
	
	This paper complements the poster \cite{ciu25b}, which was on display during the \href{https://sites.google.com/view/prin-underlandscape/workshop/workshop-the-shadows-in-the-cave}{Underlandscape Workshop} held in Pisa in January 2025.
	
	\section{Methods}
	A principled design ensures that system or software development follows well-defined theoretical, empirical, or best-practice principles to achieve consistency, robustness, and effectiveness.
	
	In our approach, the design process is guided by clearly defined principles that shape design choices and system requirements. These principles are directly influenced by the specific needs of the application environment, which, in our case, include both the unique characteristics of the touristic resource and key human factors such as user behavior, accessibility, and interaction preferences.
	
	This section delves into these aspects in detail, starting with an examination of the environmental factors that influence the design. It then proceeds to analyze the different user personas who will interact with the framework, highlighting their roles, expectations, and requirements. The framework itself consists of multiple components, with the smartphone application being a crucial element that facilitates user engagement and system functionality.
	
	\subsection{Define environment features \label{environment}}
	
	Our focus is on outdoor activities in regions that offer a rich combination of natural landscapes, human-made structures, or historically significant sites. These locations can range from archaeological sites and national parks to urban areas with cultural landmarks.
	
	These locations are often significant not only for their individual points of interest but also for the delicate balance of their broader landscape. This balance can encompass both architectural and ecological dimensions, where natural and human-made elements coexist in a way that contributes to the site's unique character. It is therefore essential that exploration and interpretation tools remain non-invasive, ensuring that the integrity of the environment is preserved. Any intervention—whether physical modifications, infrastructure, or technological aids—must be carefully designed to avoid disrupting the visual, structural, or ecological harmony of the site.
	
	In this context, the existence of predefined trails is not the primary concern; what matters most is the precise location and significance of the Points of Interest (POIs). This is particularly relevant in two types of scenarios:
	\begin{itemize}
	\item Absence of trails – Some locations, such as open archaeological sites or vast natural areas, may lack marked paths altogether. Visitors must navigate freely, relying on contextual information rather than following fixed routes.
	\item Irrelevance of trails – In urban environments, for instance, multiple routes may lead to the same POI, making any single path non-essential to the visitor experience. The journey is flexible, allowing different approaches to exploration.
	\end{itemize}
	
	Another key challenge is that identifying a POI is not always straightforward. Unlike conventional landmarks, certain points of interest may blend into the surrounding environment, making them difficult—or even impossible—to recognize without proper guidance. Consider, for example:
	\begin{itemize}
	\item A unique yet inconspicuous \textbf{graffiti} on a castle wall, which could go unnoticed without contextual knowledge, despite its historical or artistic significance..
	\item A \textbf{geologically significant rock formation} that holds scientific value but appears indistinguishable from other rocks to the untrained observer.
	\item A \textbf{peculiar plant species or vegetation pattern}, which may have ecological or historical significance but is difficult to distinguish from the surrounding flora without specialized knowledge.
	\end{itemize}
	
	In such cases, visitors must learn how to extract meaningful details from a visually complex landscape. They need to develop an expert’s perspective, gaining the ability to perceive and appreciate elements that are not immediately obvious. This requires both interpretative guidance and interactive learning tools to enhance their observational skills.
	
	Furthermore visitors could face certain risks associated with unstructured exploration. These hazards might include:
	\begin{itemize}	
	\item Unstable terrain, where loose rocks or uneven ground pose a risk of slipping or falling.
	\item Exposed passages, such as narrow ridges or steep areas, requiring careful navigation to ensure safety.
	\item Toxic plants resembling edible species, posing a risk of accidental ingestion.
	\item Seasonal insect hazards, where certain species present risks only during specific times of the year.
	\end{itemize}
	Given these considerations, it is essential to provide visitors with appropriate tools and guidance to enhance both their understanding of the site and their personal safety during exploration.
	\subsection{Understanding Personas in the Experience Ecosystem}
	Having established the characteristics of the target environment, we now turn our attention to the key personas involved in its experience, management, and preservation. Understanding the full range of personas is crucial for designing an effective and sustainable system. This study adopts an holistic pattern, considering not only the visitors but also the various individuals and groups who shape, support, and maintain the site’s functionality.
	
	To adopt a truly comprehensive perspective, it is insufficient to focus solely on the visitors as the central persona. While they are crucial to the experience, the site's functioning and long-term viability depend on other significant personas that contribute to its accessibility, interpretation, and upkeep.
	
	In this regard, we introduce the key personas that represent the primary stakeholders involved:
	\begin{itemize}
	\item The administrative entity persona, which represents those responsible for overseeing the site’s management, balancing social, cultural, and economic considerations to ensure the site remains accessible, well-maintained, and aligned with broader conservation or development policies.
	\item The expert team persona, composed of specialists from various fields—such as historians, archaeologists, ecologists, and cultural interpreters. These experts curate, develop, and present the cultural and scientific content associated with the site. Their role is to provide accurate, engaging, and meaningful narratives that enhance visitors' understanding and appreciation of the location.
	\end{itemize}
	By acknowledging these intertwined personas and their roles, we move beyond a visitor-centric approach and embrace a more comprehensive framework. This broader, holistic pattern incorporates the social, economic, and cultural dynamics that influence both the site’s management and the visitor experience, ensuring a balanced and sustainable engagement for all involved.
	
	\subsubsection*{The visitor: a grey nomad \label{visitor}}
	
	The term "grey nomad", as analytically defined in \cite{wil24a}, refers to mature travelers who explore a country with a focus on cultural enrichment and an open-minded attitude. While the popular stereotype of the "grey nomad" may be limited, the definition provided by the authors broadens the scope, allowing for a more diverse group of travelers to fit into this category.
	
	For the purpose of this study, we focus on a few key characteristics that define the expected visitor, which align with the concept of a grey nomad but are also applicable to a broader audience:
	\begin{itemize}
	\item \textit{Desire for an educational and enjoyable experience} -- The grey nomad seeks to engage with the environment in a meaningful way. They are not only looking for entertainment, but also for educational value, with a deep interest in learning about the local culture, history, and nature. This type of visitor is motivated by the opportunity to gain knowledge and have enriching experiences rather than simply passing through.
	
	\item \textit{Inclination toward exploration, not a "touch-and-to" experience} -- These visitors are not interested in quick, superficial visits. They prefer to explore places at a slower pace, taking the time to absorb their surroundings and engage with local people, traditions, and histories. Their travels are centered around immersion rather than a rushed itinerary, and they appreciate opportunities to explore in depth.
	\end{itemize}
	
	The "grey nomad" persona, in this expanded sense, represents travelers who embrace extended, unhurried journeys, often free from work commitments. Unlike conventional tourists who might focus on completing a predefined itinerary, these visitors prioritize the experience of the journey itself, seeking meaning and connection through slow, deliberate exploration.
	
	Grey nomads are active participants in their experiences, seeking deeper understanding and connection with the places they visit. Their journey is about more than just reaching a destination; it's about the personal enrichment they gain along the way. These visitors are drawn to slow tourism, a growing trend where the focus shifts from fast-paced, mass-market travel to a more reflective, thoughtful approach to exploration.
	
	This profile aligns closely with the growing movement of slow tourism, which emphasizes sustainability, local engagement, and a deeper, more personal connection to the environment. This ethos resonates with the values of the next persona we will discuss: the administration, whose role in site management reflects a similar commitment to thoughtful and sustainable engagement.

	\subsubsection*{The administative entity}
	
	In our scenario, the administrative entity plays a crucial role in the effort to valorize a heritage site that is currently appreciated only by locals. This represents a significant challenge, as the entity must strike a balance between promoting the site to a wider audience and preserving the authenticity and integrity of the location. A critical aspect of this task is to avoid the risks associated with overtourism, a phenomenon where the influx of visitors exceeds the capacity of the site, ultimately leading to its degradation. Overtourism not only harms the heritage itself but also diminishes the potential profits that the site could generate in a sustainable manner.
	
	The goal of the administrative entity is to ensure that the benefits of tourism are shared with the local community. The vision is to create a moderate flow of visitors, one that respects both the cultural significance of the heritage site and the way of life of the local population. This approach helps maintain the integrity of the site while also providing economic benefits to the community, without overwhelming the destination.
	
	Given this focus on sustainable and respectful tourism, the expected financial income from such an initiative will necessarily be limited. As a result, the funding available for promotional activities is also constrained. The administrative entity must therefore operate within a low-budget framework, prioritizing targeted, efficient promotion that attracts visitors without causing harm to the site or the surrounding community.
	
	\subsubsection*{The experts}
	While the administrative entity is responsible for defining and managing a valorization strategy, its successful implementation relies on the expertise of two distinct groups of specialists: one dedicated to developing the content and another focused on devising strategies to effectively convey it.
	
	The creation of educational content for the geotouristic experience requires a diverse range of competencies that encompass both scientific knowledge and cultural interpretation. The role of the scientific community is straightforward, as their expertise ensures accuracy and depth in presenting the heritage. However, the contribution of the local community, though less immediately apparent, is equally essential. Their involvement guarantees that the heritage’s significance is not only understood but also effectively communicated to visitors in an engaging and accessible way.
	
	As professional guides are well aware, visitor engagement often depends on storytelling techniques that make complex scientific or cultural concepts more relatable. A purely academic approach may not always resonate with a general audience, whereas compelling narratives, rooted in local traditions and perspectives, can foster a deeper connection between visitors and the heritage site. By integrating both scholarly insights and community-driven storytelling, the experience becomes both informative and engaging, striking a balance between authenticity and accessibility.
	
	The second group of specialists is responsible for deploying the technology needed to deliver content to visitors. A variety of technological solutions can be employed, including the production of on-site signage, the design and printing of brochures, the development of a dedicated website, and the implementation of e-tourism facilities.
	
	Achieving seamless integration between these two groups requires constant interaction and mediation. Bridging technological gaps is crucial to ensure that content creators and technology implementers work in harmony toward an integrated solution.
	
	The wider the technology gap, the more challenging it becomes to achieve a satisfactory integration. This highlights the importance of user-friendly technologies that facilitate collaboration. For instance, a simple, intuitive editing tool may enable content designers and typographers to work together efficiently, whereas a complex, feature-rich professional publishing software could create unnecessary barriers to communication.
	
	The remainder of this paper delves into these aspects, using the design of an e-tourism tool as a case study.
	
	\subsection{Design principles}
	Having established the application scenario and outlined general considerations regarding a valorization initiative, we now transition to defining the design principles that will guide the development of an operational prototype.
	
	To ensure a structured approach, we categorize these design principles, summarized in table \ref{tab:principles}, into three distinct groups:
	\begin{itemize}
	\item User-Centered Principles – focusing on addressing the needs, expectations, and engagement of the visitors.
	\item Management and Content Design Principles – aimed at ensuring effective administration, content organization, and seamless integration of technological solutions.
	\item Sustainability Principles - ensuring a balanced integration of the solution’s impact across multiple dimensions and its long-term value and benefits
	\end{itemize}
	
	By structuring the design process around these principles, we aim to create a balanced and sustainable framework that enhances both the visitor experience and the long-term management of the heritage site.
	
	\subsubsection*{User oriented design principles}
	In designing the prototype, a fundamental priority is ensuring an intuitive and accessible user experience. Visitors engaging with the system may have diverse levels of technological proficiency and varying expectations regarding content delivery. To accommodate these factors, we establish the following user-oriented design principles:
	
	\begin{enumerate}
	\item  \textbf{Simplicity of the User Interface (Avoiding Feature Creep)}
	\\ A minimalistic and intuitive user interface (UI) is essential to facilitate engagement without overwhelming users. Unnecessary complexity and excessive functionality ("feature creep") should be avoided, ensuring that interactions remain straightforward and purposeful. The interface should provide only the essential tools needed for navigation, information retrieval, and interaction with the environment.
	
	\item \textbf{POI Localization Relative to the User}\\
	Visitors should be able to easily determine the location of Points of Interest (POIs) in relation to their current position. The system should offer clear directional guidance, whether through a simple map, on site markers, or proximity-based alerts. This feature ensures that users can orient themselves efficiently, even in open or unstructured environments where predefined paths may not exist.
	
	\item \textbf{Visual Identification of the POI}\\
	Recognizing a POI within its natural or urban environment can be challenging. The system should provide visual cues to assist users in identifying the exact physical object or site of interest. This may involve annotated images, overlays on digital maps, or augmented reality elements, ensuring that visitors can correctly distinguish the POI from surrounding features.
	
	\item \textbf{Limited Textual Description and Accessibility}\\
	To ensure usability and prevent cognitive overload, textual descriptions should be concise and focused. While we target slow tourism, we assume that visitors will dedicate only a few minutes to reading—equivalent to a few hundred words.
	
	To accommodate visual impairments, all textual content should be convertible to speech, enabling voice-assisted navigation. Additionally, translation options should be available to ensure accessibility for international visitors. Given these requirements, text-based descriptions are the preferred choice, as they can be easily processed by offline translation tools and text-to-speech converters available on smartphones. In contrast, more complex media, such as image galleries or augmented reality content, often depend on an Internet connection and may ultimately distract visitors from the experience.
	
	\item \textbf{Offline Operation and GPS Availability}\\
	Given that many heritage sites and natural areas may have limited or no internet connectivity, the system must be designed to function offline. Users should be able to access maps, POI information, and navigation tools without requiring a network connection. Additionally, GPS signals may be unreliable in areas with dense buildings, remote locations, or obstructions such as forests. In such cases, alternative methods for determining the user’s location should be available, such as manual location input or proximity-based cues using nearby landmarks. This ensures that the system remains functional and reliable even when internet or GPS signals are unavailable.
	
	\end{enumerate}
	
	By adhering to these principles, the design aims to provide an efficient, engaging, and accessible experience, ensuring that visitors can seamlessly explore and appreciate the site without unnecessary technological barriers.
	
	\subsubsection*{Stakeholder-Oriented Design Principles}
	The success of the valorization initiative and the sustainability of the project depend not only on the user experience but also on how well the system supports the interests and operational needs of the various stakeholders involved. For this reason, we establish the following stakeholder-oriented design principles:
	
	\begin{enumerate}
	
	\item \textbf{Cost-Effective Solutions for Financial Constraints}\\
	Due to the expected limited income from the initiative, it is crucial that the design of both the technological and content-related aspects is done with an eye on cost-efficiency. Financial constraints mean that solutions must be affordable to implement and maintain over time. The system should prioritize low-cost technologies, open-source software, and scalable solutions to ensure that the initiative remains within budget while still delivering a high-quality experience. Additionally, any financial investments should be modular and flexible, allowing for gradual improvements rather than requiring large, upfront costs.
	
	\item \textbf{Reusable and Shareable for Economic Efficiency}\\
	To maximize the economic sustainability of the initiative, it is essential that the resources and tools developed for the project are reusable and shareable. This principle ensures that the system can be adapted and reused across different contexts, enabling cost efficiency and broader impact. The content created should be modular, allowing for its reuse in different locations or for different purposes, thus avoiding the need for constant redevelopment. Additionally, the sharing of content and tools between stakeholders can reduce costs and support collaboration among regional or national entities focused on heritage preservation and tourism.
	
	\item \textbf{Open Content Format for Long-Term Sustainability}\\
	To ensure the long-term sustainability of the content, it must be designed using open formats. Open formats allow content to be easily accessed, updated, and shared, even as technologies evolve or stakeholders change over time. The use of standardized, open-source tools and formats makes it easier for different groups to collaborate, as it ensures that the content remains flexible and adaptable to future developments. Moreover, it supports the preservation of the heritage experience, ensuring that the project can continue to be used and updated without being dependent on proprietary systems or vendors.
	
	\item \textbf{User-Friendly Interface for Content Definition}\\
	The system should provide user-friendly tools for content creation and management, particularly for the stakeholders involved in the development and maintenance of the educational material. These tools should be accessible even to those with limited technical expertise, facilitating collaboration among experts from various fields. By simplifying the content creation process, the system will allow specialists in heritage management, education, and local history to collaborate more effectively. The design of these tools should prioritize ease of use, enabling content contributors to focus on the subject matter, rather than getting bogged down by complex technical issues.
	\end{enumerate}
	
	\subsubsection*{Sustainability-Oriented Design Principles}
	In line with the long-term goals of the project and its environmental and financial sustainability, the following principles guide the design to minimize resource consumption and ensure the initiative’s continued viability:
	
	\begin{enumerate}

	\item \textbf{Utilization of Public Web Services Over Private Resources}\\
	To ensure the sustainability and accessibility of the project in the long term, it is crucial to rely on public web services rather than private or proprietary resources. Public services tend to be more stable, open, and affordable, providing the necessary long-term availability and support. This reduces the risk of dependency on a single vendor and mitigates potential costs associated with licensing or service discontinuation. Furthermore, public resources are typically aligned with open standards, which supports interoperability with other tools and systems, ensuring greater flexibility for future technological integrations.
	
	\begin{figure}
		\centering
		\includegraphics[width=0.9\linewidth]{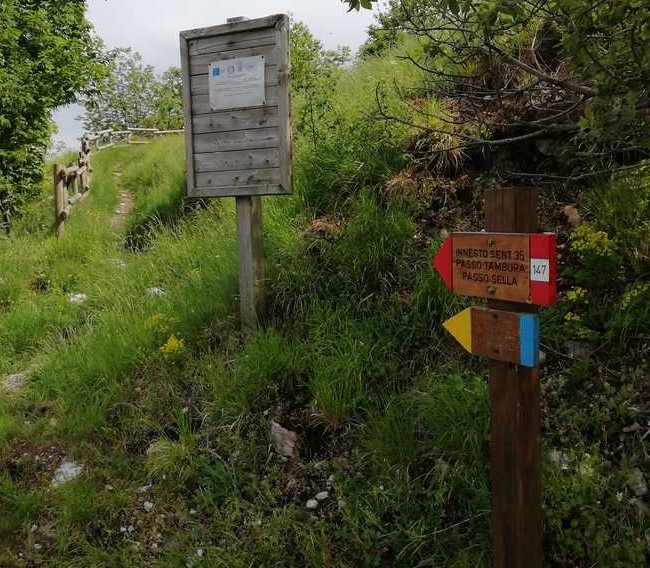}
		\caption[An example of traditional signage]{An example of traditional signage. One of the recent panels provides useful information, while the meaning of the other remains unclear. On the left, an older panel in good condition—likely replaced by the new one—features a pinned post that is now unreadable.}
		\label{fig:traditional}
	\end{figure}
	
	\item \textbf{Limited Impact of On-Site Signage Devices}\\
	On-site signage is an important part of the visitor experience, but it also comes with sustainability challenges. Traditional signage can become obsolete, damaged, or outdated over time, requiring constant maintenance or replacement, which involves both material and financial costs (see Figure \ref*{fig:traditional}). To mitigate this impact, the project should prioritize minimal and flexible signage solutions that are easy to update. This could include digital signage that can be remotely updated to reflect changing information or events. By keeping the physical signage footprint limited and modular, the initiative can significantly reduce waste and the ongoing costs associated with on-site updates, while ensuring that the visitor experience remains current and relevant.
	
	\item \textbf{Ensuring the Financial Sustainability of On-Site Signage}\\Long-term financial sustainability of on-site signage must be carefully considered, as continuous investment in signage maintenance can quickly become a financial burden. Emphasizing the temporary nature of some signage and combining it with digital alternatives or mobile solutions (e.g., smartphone apps) can reduce the financial strain and allow for more cost-effective upkeep.
	
	\end{enumerate}

	\section{Results}
	
The design principles outlined above emerged from my experience as a computer scientist within a multidisciplinary team. Throughout this process, I collaborated closely with experts from various fields, and ongoing discussions with non-computer science participants were instrumental in refining and shaping the guiding design principles. The app described in the following section is a direct attempt to apply these principles, offering visitors a structured way to explore a small historic area in Tuscany, home to a series of abandoned mills known mainly to a small group of experts and local residents.

The case study focuses on a small region in Tuscany, north of Lucca, where the remains of eight hydraulic mills are concentrated within an area of approximately one square kilometer near the village of Montefegatesi. These mills represent an important historical heritage, shedding light on a lost yet sophisticated technology. Although the site is located near Bagni di Lucca, a well-known tourist destination, it remains largely unrecognized and lacks proper maintenance. Existing trails, where present, are in poor condition, making it a fitting example of the type of location described in Section \ref{environment}.

Montefegatesi, a village with fewer than 200 inhabitants, is accessible from Bagni di Lucca via a 14 km mountain road requiring a 30-minute drive. This accessibility constraint naturally positions visitors as slow tourists, willing to dedicate a full day to the journey, including the village itself and the nearby mills. This aligns with the user profile described in Section \ref{visitor}.

The administrative stakeholders include the municipality of Bagni di Lucca and the local community. The municipality considers Montefegatesi a peripheral village and reportedly neglects road maintenance (see the \href{ https://www.facebook.com/story.php?story_fbid=1097509845107985&id=100045469400975&_rdr}{Facebook Story}). Meanwhile, the locals actively manage Facebook and Instagram pages featuring community announcements, reflecting a proud and engaged population.

Among the experts supporting the valorization efforts, we include Maria Pardini, a master's student in Digital Humanities, who dedicated her thesis \cite{par24a} to documenting the site and developing the first version of the App described in the next section. As part of her research, Pardini organized a reconnaissance trip to Montefegatesi, accompanied by Graziano Serafini, a local expert author of a monography on the topic \cite{ser21a}, and Prof. Enrica Salvatori, who provided cultural and historical insights. I was responsible for supervising the design of the \textit{Turista} App within the framework of Pardini's thesis and personally carried out its subsequent re-engineering.

The case study presented here aligns well with our initial premises—unsurprisingly, given that the work is part of \href{https://sites.google.com/view/prin-underlandscape/progetto}{Underlandscape}, an Italian research initiative classified as a \textit{Research Project of National Interest} (PRIN). One of the project's key objectives is to develop strategies for promoting sustainable tourism in pristine natural sites.

In table \ref{tab:principles} we include the identified design principles with the abbreviations used in the paper.

The next section presents the revised version of the Android App, developed as an application of the design principles to the defined use case.

\begin{table}
	\scriptsize
	\begin{tabular}{|p{65mm}|l|}
		\hline
		\multicolumn{2}{|l|}{\textbf{User Oriented design principles}}\\
		\hline
		Simplicity of the User Interface & U1 \\
		\hline
		POI Localization Relative to the User & U2	\\
		\hline
		Visual Identification of the POI & U3 \\
		\hline
		Limited Textual Description & U4\\
		\hline
		Offline Operation and GPS Availability & U5 \\
		\hline 
		\multicolumn{2}{|l|}{\textbf{Stakeholder Oriented design principles}}\\
		\hline
		Cost-Effective Solutions and Financial Constraints & St1 \\
		\hline
		Reusable and Shareable for Economic Efficiency & St2 \\
		\hline
		Open Content Format for Long-Term Sustainability & St3 \\
		\hline
		User-Friendly Interface for Content Definition & St4\\
		\hline 
		\multicolumn{2}{|l|}{\textbf{Sustainability Oriented design principles}}\\
		\hline
		Utilization of Public Web Services Over Private Resources & Sus1\\
		\hline
		Limited Impact of On-Site Signage Devices & Sus2\\
		\hline
	\end{tabular}
	\caption{Design principles by name and code \label{tab:principles}}
\end{table}

\subsection{Implementation of the design principles}

We choose a solution based on a standalone smartphone app with no deployment infrastructure, as the user carries all the necessary hardware. This approach, in strict adherence to St1, acknowledges the project's financial constraints—relying solely on individual donations—and its dependence on volunteer-driven management. In the discussion section, we will explore a less restrictive alternative.

This smartphone-only approach also aligns with the Sus2 design principle by minimizing environmental impact, as it eliminates the need for signage or any on-site infrastructure.

Given the limited programming skills of our student, whose background is in Digital Humanities, I recommended the AppInventor framework \cite{pat19a} as the development tool. This platform provides a visual editor and an intuitive development environment, making it suitable for non-professional coders. This choice ultimately aligns with principle St4, as it enables the convergence of two skill sets within the same individual.

The resulting application, however, was not entirely satisfactory. In particular, the content was hardcoded within the app, preventing reuse without modifying the source code—thus violating the St2 principle.

Nevertheless, the student's implementation provided valuable insights that informed the subsequent re-engineering, which retained the AppInventor framework for development.

\subsubsection*{An all-inclusive configuration file}

In the revised version, the App functions as a content-less skeleton framework. The location-specific content is loaded when the application is launched for the first time. Subsequent uses rely on the same content, meaning that the user must reinstall the application to access different content.

This solution represents a tradeoff with respect to principle U1: while the initial setup is a one-time operation, reconfiguration is more complex. However, this aligns with the intended single-use nature of the App, as users are expected to uninstall it after their visit.

Loading the configuration file requires an Internet connection. However, since the configuration is stored within the App, subsequent restarts do not rely on connectivity, thereby supporting principle U5. We will further discuss this aspect in section \ref{sec:discussion}.

To simplify the acquisition of the configuration file’s URL, we adopt QR code technology. Upon first launch, the App prompts users to scan a QR code, which can be found on a brochure or at the site. Alternatively, users can manually enter the URL after copying it from a document on their smartphone. This setup process is designed to be straightforward, in line with design principle U1.

\subsubsection*{A Focused Single-Page Layout}

The user interface presents all relevant information on a single page.

\begin{figure}
	\textcolor{purple}{
		\setlength{\fboxsep}{0pt}
		\setlength{\fboxrule}{5pt}
		\centering
		\fbox{\includegraphics[width=0.9\linewidth]{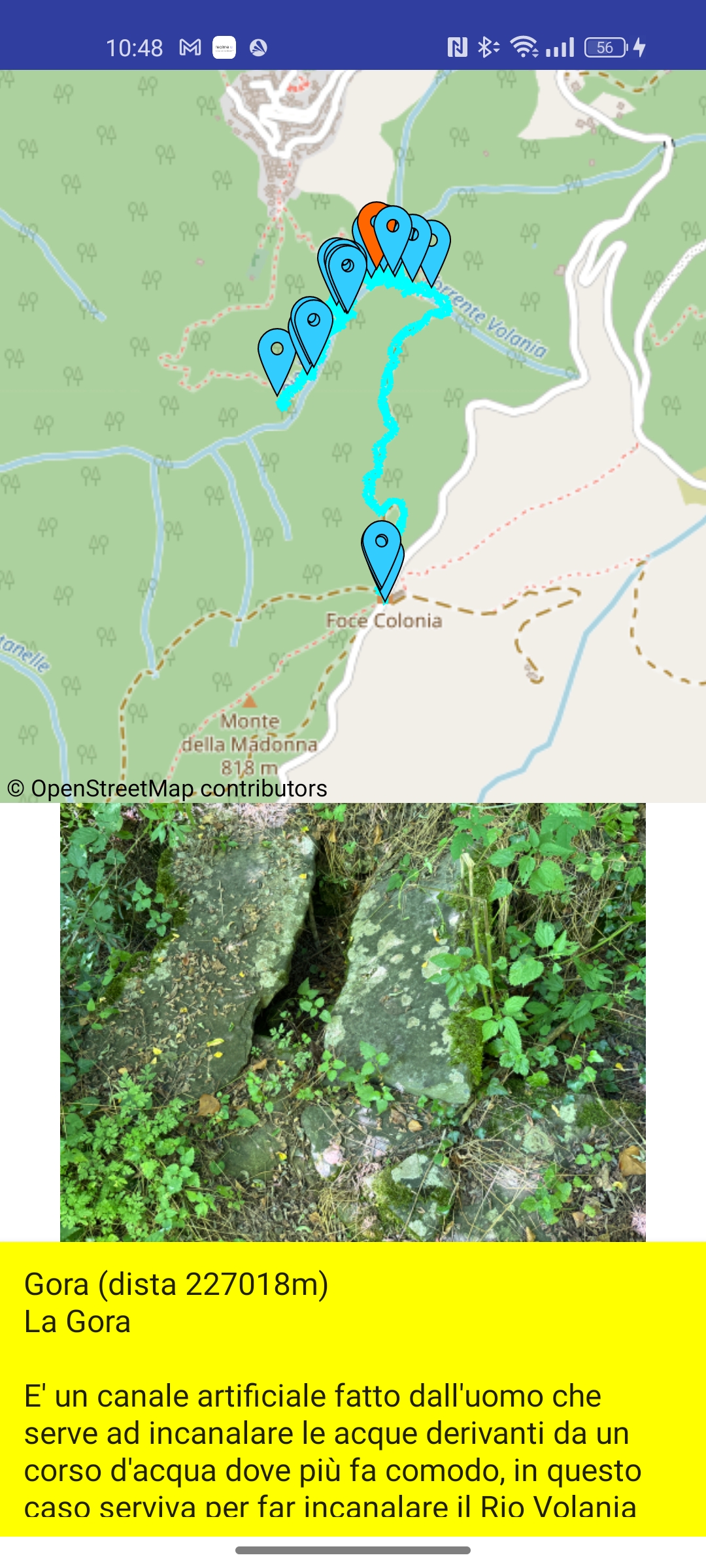}}
	}
	\caption[App screen]{The layout of the app interface. At the top, the map displays all POIs, the user's position, and the nearest highlighted POI. In the middle, an image of the selected POI is shown. The bottom section contains its description, possibly truncated to fit the screen. The map can be navigated using standard gestures, and tapping on a POI reveals its title. The image can be enlarged by tapping on it, and similarly, the full description can be accessed if truncated.}
	\label{fig:3parti}
\end{figure}

The top half of the screen displays a map showing the user's current position (obtained via the smartphone's GPS receiver) along with nearby points of interest (POIs). The user can interact with the map using standard gestures, such as zooming and panning. Tapping on a POI reveals its name, while the nearest POI is highlighted in a different color for easy identification. In addition to those already present on the base map (sourced from OpenStreetMap), the configuration may include a recommended path for navigation.

The lower half of the screen provides details about the nearest POI. It includes an image to aid visual identification and a brief textual description that specifies the POI’s distance from the user’s GPS position and the beginning of a longer description.

The user can enlarge either the image or the text with a simple tap, switching to a full-screen view. Tapping the screen again restores the original three-part layout.

No additional features are provided. The intended workflow assumes that the user accesses the installation link and QR code for setup via the location's website while planning their visit. Once on-site, the visitor selects a POI to explore and identifies it by name. Navigation to the POI can be guided either by a suggested trail displayed on the map or by existing ground tracks. Upon arrival, the visitor confirms the POI using the image for visual identification and reads the accompanying description by tapping the text frame in the app. This process is then repeated for the next POI.

Despite its minimalistic design, the interface meets essential usability requirements: it provides guidance (U2 principle) and identification (U3 principle). The textual description is limited to a few hundred words but can be expanded by scrolling. Additionally, users can copy and paste the text into an offline translator for language conversion or into a text-to-speech converter for auditory assistance, reinforcing compliance with the U4 principle.

\subsection{Easy management of the configuration file}

\begin{figure*}
\begin{lstlisting}[language=json]
{
  "type": "FeatureCollection",
  "properties": {
    "Nome": "Mulini_Montefegatesi"
  },
  "features": [
    {
      "type": "Feature",
      "geometry": {
        "type": "Point",
        "coordinates": [10.594269,44.058392]
      },
      "properties": {
        "ulsp_type": "POI",
        "Titolo": "Elemento terminale di una tromba di carico",
        "Descrizione": "Aiutava a indirizzare l'acqua alle pale della turbina il legno che a sua volta faceva azionare il movimento della macina."
      }
    }
  ]
}
\end{lstlisting}
	\caption{Configuration file example with a single POI. The complete file for the use case is available at  \href{https://raw.githubusercontent.com/prin-underlandscape/Itinerario_Mulini_Montefegatesi/refs/heads/main/Itinerario_Mulini_Montefegatesi.geojson}{this link}, selecting the \textit{raw} option using the button on the left side of the GitHub page \label{code:configuration}}.
\end{figure*}

The configuration file is a human-readable GeoJSON file containing a single FeatureCollection object, which is one of the GeoJSON types  \cite{rfc7946}. The GeoJSON format provides the "properties" property, associated to a JSON object, to customize the content of a feature, or of the whole FeatureCollection. The content of the configuration file for our app is described by some well-defined properties included in the properties object of POIs, or of the whole FeatureCollection. See an example in figure \ref{code:configuration}. The description of the defined properties is in JSON file that is available to the public, located at \href{https://raw.githubusercontent.com/prin-underlandscape/OfflineForm/refs/heads/main/formats.js}{this link}, for compliance with the design principle St3. The file has been developed following a defined workflow which is documented in \cite{ciu25c}.

Although the configuration file can be edited using a generic text editor, a graphical interface is more suitable for non-coders.

\begin{figure*}[t]
	\centering
	\includegraphics[width=0.7\linewidth]{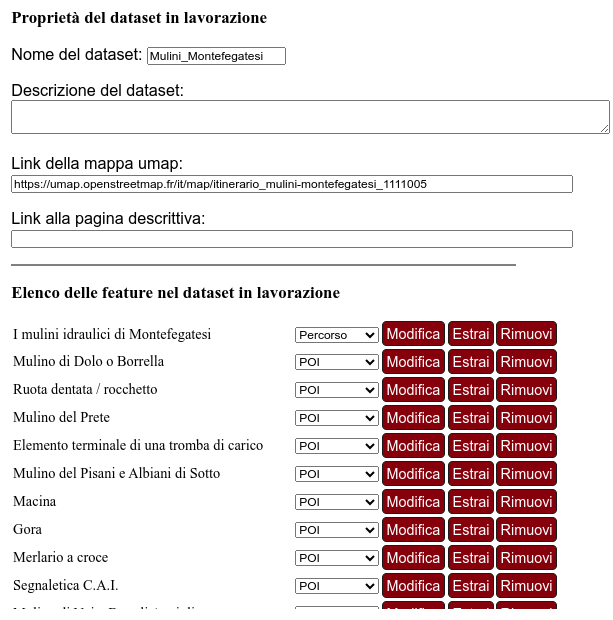}
	\caption[Off Form]{The \textit{Off} form to edit dataset contents}
	\label{fig:off}
\end{figure*}

\begin{figure*}[t]
	\centering
	\includegraphics[width=0.7\linewidth]{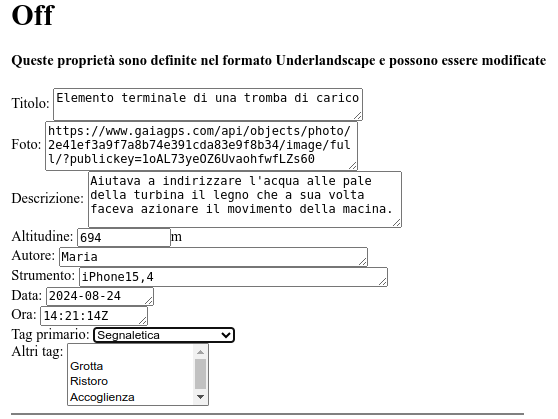}
	\caption[Form-2]{The \textit{Off} form to edit a POI feature contents}
	\label{fig:poiform}
\end{figure*}

To facilitate the creation of a well-formed configuration file, we developed \textit{Off} (Offline Formatter), a JavaScript-based tool that provides a form-like interface for editing. Since it is written in JavaScript, \textit{Off} is fully portable and can operate offline, making it accessible in various environments. Figures \ref{fig:off} and \ref{fig:poiform} illustrate its functionality: the former displays an overview of the configuration file with its attributes and a portion of the POI list, while the latter shows the POI editing form.

Figures \ref{fig:off} and \ref{fig:poiform} illustrate the operation of this JavaScript tool. Figure \ref{fig:off} presents an overview of the entire configuration file as displayed by the tool, showing its attributes along with a portion of the POI list. Figure \ref{fig:poiform} shows the POI editing form, where users can modify individual attributes through a structured interface.

The configuration file must be accessible via a web URL, requiring cloud storage support. While there are no restrictions on the choice of web storage, we opted for a GitHub repository. This solution is preferable to hosting it on a website, as the latter would likely be managed by an unstructured group of volunteers, leading to frequent redesigns and potential disruptions as responsibilities shift. In contrast, a GitHub repository serves a focused purpose and remains stable, unaffected by unplanned modifications.

By using a public web service like GitHub, this approach aligns with the Sus1 principle, which advocates for the use of public web services over on-premises resources. This choice ensures the sustainability of the service without the need for maintaining physical infrastructure or managing complex server setups.

The repository for our use case is available at \href{https://github.com/prin-underlandscape/Itinerario_Mulini_Montefegatesi}{this link}. Additionally, we have also hosted on GitHub the repository for the \textit{Off} script (\href{https://github.com/prin-underlandscape/OfflineForm}{here}), enabling it to run directly from a GitHub page, as well as the Android app \textit{Turista} (\href{https://github.com/prin-underlandscape/AppTurista}{here}).

In summary, the GitHub-based solution supports the Sus1 principle ensuring long-term sustainability without requiring ongoing management.

\section{Discussion \label{sec:discussion}}

Many smartphone apps assist users navigating wilderness areas. These apps can generally be categorized into two types:

\begin{itemize}
	\item \textbf{Active tools} – users can create content.
	\item \textbf{Passive tools} – users cannot create new content.
\end{itemize}

\textbf{Active tools} rely on a complex infrastructure that enables users to generate geo-referenced data along their route and export it to an external platform for further editing, sharing, and documentation. Examples of such tools include GeoPaparazzi \cite{ded10a,pap22a}, GaiaGPS, and Locus Map. While these applications provide numerous features to support content creation, many of these functionalities are unnecessary for our purposes and conflict with our design principle U1. Consequently, we do not consider them in our study.

\textbf{Passive tools} are often studied in the context of Virtual Reality (VR), as documented in the survey paper \cite{kek17a}. However, VR remains an expensive technology, and we do not expect visitors within our target profile to carry such equipment, nor do we anticipate local communities to provide it. When excluding advanced and costly technologies, the development of a navigation app is generally not considered a significant research challenge. Existing solutions in this category include izi.travel (\href{https://izi.travel}{link}), which supports the creation and use of browser-based guided audio tours.

The app introduced in this paper belongs to the second category, prioritizing simplicity over advanced features. The added value of our research lies in integrating the app’s design within a comprehensive framework that encompasses content production and aligns with specific design principles derived from the expectations of various stakeholders. This holistic approach is crucial for successful deployment but is rarely addressed in existing literature on the subject.

The app relies on the user's location, which in our case depends on the availability of GPS signals. However, GPS may sometimes be insufficient for accurate positioning or may provide incorrect locations. This issue becomes particularly significant in small-scale applications, such as walking navigation, where dense vegetation can further interfere with GPS accuracy. Alternative technologies exist, as discussed in \cite{kee19a}

QR markers placed at key locations may provide the required localization. When a user scans a QR marker, the app receives precise location data and can provide appropriate guidance. One minor drawback of QR markers is the initial effort required for their production and placement. While this entails some investment, the cost is considered acceptable due to its limited impact on overall performance regarding Sus2 and St1.

We implemented such a solution in a different tool which is not directly part of the current application \cite{der23a}, but the results obtained are highly relevant, as the challenges and limitations identified apply equally to this case.

Specifically, we found that although QR markers provide precise localization when scanned, their long-term viability is hindered by maintenance issues. Even when made from durable plastic, they are susceptible to removal or damage due to weather conditions or human interference. Therefore, ensuring their continued availability requires on-site inspections and replacements, making management a complex and resource-intensive task. These findings led us to conclude that, despite the potential benefits of QR-based localization, its drawbacks make it less suitable for our prototype. 

After careful evaluation, we ultimately opted for a GPS-based approach for our prototype.

Another alternative we considered is the use of passive RFID tags instead of QR codes. While RFID tags offer slightly better resistance to environmental conditions, the smartphone-compatible 13.56 MHz technology has a maximum operational range of about 1.5 meters—reduced even further when using a phone as the receiver. As a result, the user must locate and physically approach the tag to scan it, offering little advantage over QR markers.

A more viable alternative is the use of Bluetooth beacons. These devices typically have an operating range of 2 to 5 meters, comparable an optimal GPS accuracy. Additionally, beacons can be encased in protective enclosures to prevent deterioration. While the cost per unit is approximately 20 euros—a reasonable expense even for small communities—beacons require periodic maintenance to replace depleted batteries.
	
\section{Conclusion}

Technology provides us with new tools every day to assist in our activities. As a result, selecting the most effective tools while excluding unnecessary ones has become increasingly challenging. A clear statement of design principles—derived from application requirements and the defined operational environment—serves as a crucial criterion for making these decisions.

We applied these principles to the design of a framework aimed at the sustainable valorization of a region with significant tourist potential. Our analysis led us to adopt the motto: "the simpler, the better." While advanced technologies such as augmented reality offer sophisticated interfaces, they do not address the primary need for accurate user localization. The framework consists of a content format, tools for content creation and editing, and an Android app for user guidance. We designed a prototype that relies on a straightforward localization solution—the Global Positioning System (GPS)—and on-field trials confirmed its usability, provided that the GPS maintains nominal accuracy.

We also explored alternative technologies, such as QR tags and microwave beacons. One viable but somewhat controversial option is Bluetooth beacons, which could enable user localization; however, they present challenges similar to those of QR codes and come at a higher cost.

Ultimately, our approach envisions integrating these technologies. However, this integration must remain seamless from the user's perspective, ensuring a simple and immediately usable interface.

	\bibliographystyle{plainurl} 
	\bibliography{paper} 

\begin{thebibliography}{10}

\bibitem{rfc7946}
H.~Butler, M.~Daly, A.~Doyle, Sean Gillies, T.~Schaub, and Stefan Hagen.
\newblock {The GeoJSON Format}.
\newblock RFC 7946, August 2016.
\newblock URL: \url{https://www.rfc-editor.org/info/rfc7946}, \href
  {https://doi.org/10.17487/RFC7946} {\path{doi:10.17487/RFC7946}}.

\bibitem{ciu25c}
A.~Ciuffoletti and L.~Chiti.
\newblock Sustainable open-data management for field research: A cloud-based
  approach in the underlandscape project, 2025.
\newblock \href {https://doi.org/http://dx.doi.org/10.13140/RG.2.2.13940.05761}
  {\path{doi:http://dx.doi.org/10.13140/RG.2.2.13940.05761}}.

\bibitem{ciu25b}
Augusto Ciuffoletti.
\newblock App android per una visita guidata - un approccio orientato ad
  estensibilità e semplicità d'uso, 2025.
\newblock \href {https://doi.org/http://dx.doi.org/10.13140/RG.2.2.13224.89601}
  {\path{doi:http://dx.doi.org/10.13140/RG.2.2.13224.89601}}.

\bibitem{ded10a}
Mauro De~Donatis, Andrea Antonello, Luca Lanteri, Sara Susini, and Marco Foi.
\newblock Beegis: a new open source and multiplatform mobile gis.
\newblock {\em US Geological Survey Open-File Report}, 2010:1335, 2010.

\bibitem{der23a}
Maria~Grazia Deri, Letizia Chiti, and Augusto Ciuffoletti.
\newblock A sustainable approach to tourist signage on heritage trails.
\newblock {\em Sustainability}, 15(23):16251, 2023.

\bibitem{kek17a}
Anabel~L Ke{\v{c}}ke{\v{s}} and Igor Tomi{\v{c}}i{\'c}.
\newblock Augmented reality in tourism - research and applications overview.
\newblock {\em Interdisciplinary Description of Complex Systems}, 15:157--167,
  2017.
\newblock URL: \url{https://api.semanticscholar.org/CorpusID:11752018}.

\bibitem{kee19a}
Robert~F Keefe, Ann~M Wempe, Ryer~M Becker, Eloise~G Zimbelman, Emily~S Nagler,
  Sophie~L Gilbert, and Christopher~C Caudill.
\newblock Positioning methods and the use of location and activity data in
  forests.
\newblock {\em Forests}, 10(5):458, 2019.

\bibitem{pap22a}
Elisa Paperini, Francesca Anichini, Gabriele Gattiglia, et~al.
\newblock “field data collection app” e ricognizioni archeologiche:
  Geopaparazzi.
\newblock {\em Archeologia e Calcolatori}, 33(2):215--234, 2022.

\bibitem{par24a}
Maria Pardini.
\newblock Turista - applicazione che mira a rendere il patrimonio culturale
  accessibile, velocemente a tutti.
\newblock Bachelor thesis, University of Pisa, 2024.

\bibitem{pat19a}
Evan~W Patton, Michael Tissenbaum, and Farzeen Harunani.
\newblock Mit app inventor: Objectives, design, and development.
\newblock In {\em Computational thinking education}, pages 31--49. Springer
  Singapore Singapore, 2019.

\bibitem{ser21a}
Graziano Serafini.
\newblock {\em I mulini idraulici nel comune di Bagni di Lucca. Visita guidata
  agli 84 vecchi mulini}.
\newblock 2021.

\bibitem{wil24a}
Mark~A Williams, Xinyuan Wang, Melinda~T McHenry, and Angus~M Robinson.
\newblock Australian geotourism discovery platform: A sustainable and
  user-friendly platform for accessible exploration of geosites, geotrails,
  cultural, and mining heritage sites.
\newblock {\em Sustainability}, 16(13):5482, 2024.
\newblock User personas.

\end{thebibliography}
	
\end{document}